\documentstyle{mn}
\newif\ifAMStwofonts
\def\overleftrightarrow{\mathpalette\overleftrightarrow@}
\def\overleftrightarrow@#1#2{\vbox{\ialign{##\crcr\leftrightarrowfill@#1\crcr
 \noalign{\kern-\ex@\nointerlineskip}$\m@th\hfil#1#2\hfil$\crcr}}}
\input psfig.tex
\ifoldfss
  \ifCUPmtlplainloaded \else
    \NewTextAlphabet{textbfit} {cmbxti10} {}
    \NewTextAlphabet{textbfss} {cmssbx10} {}
    \NewMathAlphabet{mathbfit} {cmbxti10} {} % for math mode
    \NewMathAlphabet{mathbfss} {cmssbx10} {} %  "   "    "
  \fi
  \ifAMStwofonts
    \ifCUPmtlplainloaded \else
      \NewSymbolFont{upmath} {eurm10}
      \NewSymbolFont{AMSa} {msam10}
      \NewMathSymbol{\upi}     {0}{upmath}{19}
      \NewMathSymbol{\umu}     {0}{upmath}{16}
      \NewMathSymbol{\upartial}{0}{upmath}{40}
      \NewMathSymbol{\leqslant}{3}{AMSa}{36}
      \NewMathSymbol{\geqslant}{3}{AMSa}{3E}

    \fi
  \fi
\fi % End of OFSS
\ifnfssone
  \newmathalphabet{\mathit}
  \addtoversion{normal}{\mathit}{cmr}{m}{it}
  \addtoversion{bold}{\mathit}{cmr}{bx}{it}
  \newmathalphabet{\mathbfit} % math mode version of \textbfit{..}
  \addtoversion{normal}{\mathbfit}{cmr}{bx}{it}
  \addtoversion{bold}{\mathbfit}{cmr}{bx}{it}
  \newmathalphabet{\mathbfss} % math mode version of \textbfss{..}
  \addtoversion{normal}{\mathbfss}{cmss}{bx}{n}
  \addtoversion{bold}{\mathbfss}{cmss}{bx}{n}
  \ifAMStwofonts
    \ifCUPmtlplainloaded \else
      \UseAMStwoboldmath
      \makeatletter
      \new@mathgroup\upmath@group
      \define@mathgroup\mv@normal\upmath@group{eur}{m}{n}
      \define@mathgroup\mv@bold\upmath@group{eur}{b}{n}
      \edef\UPM{\hexnumber\upmath@group}
      \new@mathgroup\amsa@group
      \define@mathgroup\mv@normal\amsa@group{msa}{m}{n}
      \define@mathgroup\mv@bold\amsa@group{msa}{m}{n}
      \edef\AMSa{\hexnumber\amsa@group}
      \makeatother
      \mathchardef\upi="0\UPM19
      \mathchardef\umu="0\UPM16
      \mathchardef\upartial="0\UPM40
      \mathchardef\leqslant="3\AMSa36
      \mathchardef\geqslant="3\AMSa3E
    \fi
  \fi
\fi % End of NFSS release 1

\ifnfsstwo
  \DeclareMathAlphabet{\mathbfit}{OT1}{cmr}{bx}{it}
  \SetMathAlphabet\mathbfit{bold}{OT1}{cmr}{bx}{it}
  \DeclareMathAlphabet{\mathbfss}{OT1}{cmss}{bx}{n}
  \SetMathAlphabet\mathbfss{bold}{OT1}{cmss}{bx}{n}
  \ifAMStwofonts
    \ifCUPmtlplainloaded \else
      \DeclareSymbolFont{UPM}{U}{eur}{m}{n}
      \SetSymbolFont{UPM}{bold}{U}{eur}{b}{n}
      \DeclareSymbolFont{AMSa}{U}{msa}{m}{n}
      \DeclareMathSymbol{\upi}{0}{UPM}{"19}
      \DeclareMathSymbol{\umu}{0}{UPM}{"16}
      \DeclareMathSymbol{\upartial}{0}{UPM}{"40}
      \DeclareMathSymbol{\leqslant}{3}{AMSa}{"36}
      \DeclareMathSymbol{\geqslant}{3}{AMSa}{"3E}
    \fi
  \fi
\fi % End of NFSS release 2

\ifCUPmtlplainloaded \else
  \ifAMStwofonts \else % If no AMS fonts
    \def\upi{\pi}
    \def\umu{\mu}
    \def\upartial{\partial}
  \fi
\fi

\def\lsim{\lower.5ex\hbox{$\; \buildrel < \over \sim \;$}}
\def\gsim{\lower.5ex\hbox{$\; \buildrel > \over \sim \;$}}
\title{Computation of Mass Outflow Rate from Relativistic Quasi-Spherical 
Accretion onto Black Holes}
\author[Tapas K. Das]
{Tapas K. Das\\
Theoretical Astrophysics Group \\
 S. N. Bose National Centre For Basic Sciences \\
 Block JD Sector III Salt Lake Calcutta 700 091 India\\ 
 E-mail tdas@boson.bose.res.in}
\date{Accepted 08.04.99  Received 07.12.98. ; in original form 07.12.98}

\begin{document}
%\twocolumn
\maketitle
\begin{abstract}
We compute mass outflow rate $R_{\dot m}$ from relativistic matter
accreting quasi-spherically
onto  Schwarzschild black holes. Taking the pair-plasma pressure mediated shock
surface as the {\it effective} boundary layer (of the black hole) from where
bulk of the outflow is assumed to be generated, computation of this rate is
done using combinations of exact transonic inflow and outflow solutions.
We find that $R_{\dot m}$ depends on the initial parameters of the flow,
the polytropic index of matter, the degree of compression of matter
near the shock surface and on the location of the shock surface itself.
We thus not only study the variation of the mass outflow
rate as a function of various physical parameters governing the problem
but also provide a sufficiently plausible estimation of this rate.
\end{abstract}

\noindent
\begin{keywords}
AGN -- quasars -- jets -- accretion, accretion discs --  black hole physics -- shock-waves -- hydrodynamics 
-- outflow -- wind \\ \\
{\bf To appear in MNRAS}   

\end{keywords}
% if the twocol comand is here then two col after abstract, i,e, from 
%2nd page
%\twocolumn
\section{Introduction}
\noindent

Computation of mass outflow rates from the advective accretion 
disks around black holes and neutron stars very recently has been done
(Chakrabarti, 1998; Das, 1998, 1998a; Das \& Chakrabarti, 1998 [DC98 hereafter]).
by self consistently combining the exact transonic accretion and wind solutions.
Rigoriously justifying the fact that most of the outflowing matter 
comes out from the CENtrifugal
Pressure Supported BOundary Layer (CENBOL)
it has been shown there that the 
Rankine-Hugoniot shock location or the location of the maximum polytropic
pressure acts as the CENBOL. However, for some black hole models of active galactic 
nuclei, inflow may not have accretion disk (Rees, 1977 and references therein).
Accretion
is then quasi-spherical having almost zero or negligible angular 
momentum (Bondi [1952] type accretion) and the shock is not of 
Rankine-Hugoniot
type. On the 
otherhand, absence of angular momentum rules out the possibilty of 
formation of the polytropic pressure maxima. So for quasi-spherical accretion,
absence of intrinsic angular momentum of the accreting material
does not allow CENBOL formation. 
It has been shown that (Meszaros \& Ostriker, 1983; Kazanas \& Ellison, 1986 [KE86
hereafter])
for quasi -spherical 
accretion onto 
black holes, steady state situation may be developed where a standing 
collisionless shock may form due to the plasma instabilities and for
nonlinearity introduced by small density perturbation. 
This is because, 
after crossing the sonic point
 the infalling matter (in plasma form) becomes highly supersonic.
Any small perturbation and slowing down of the infall velocity
will create a piston and produce a shock. A spherically symmetric shock 
produced in such a way will accelerate a fraction of the inflowing
plasma to relativistic energies. The shock accelereted 
relativistic particles 
suffer essentially no Compton loss and are assumed to lose energy
only through $p - p$ collision. 
These relativistic hadrons are not readily captured by the 
black hole (Protheros \& Kazanas, 1983) rather
considereble high energy density of these
relativistic protons would be maintained to support a standing, collisionless,
spherical shock around the black hole (see KE86  and references therein).
Thus a self-supported standing shock may be produced {\it even} for accretion
with zero angular momentum.
In this work, we take this 
pair-plasma pressure mediated shock surface as the alternative
of the CENBOL which can be treated as the {\it effective} physical hard
surface which, in principle  mimics the ordinary stellar surface regarding 
the mass outflow.
The condition necessery for the development and maintainence of such a self
supported spherical shock is 
satisfied for the high Mach number solutions (Ellison \& Eichler, 1984).
Keeping this in the back of our mind, for our present work,
we concentrate only on low energy accretion to obtain high shock Mach number.
Considering low energy 
(${{\cal E} \lsim  0.001}$) accretion, 
we assume that particles accreting
toward black hole are shock accelerated via first order Fermi
acceleration (Axford, et. all, 1977)
producing relativistic protons. Those relativistic protons
usually scatter several times before being captured by the black hole.
These energised particles, in turn, provide sufficiently outward
pressure to support a standing,
collisionless shock. A fraction of the energy flux of infalling matter
is assumed to be converted into radiation at the shock standoff
distance through hadronic ($p -p$) collision and mesonic ($\pi^{\pm},\pi^0$) 
decay. Luminosity produced by this fraction is used to obtain the shock 
location for the present work. Our approach to formulate the expression
for
shock location is somewhat similar to that of KE86.\\
At the shock surface,
density of the post-shock material shoots up and velocity falls
down, infalling matter starts piling up on the shock surface. The post
shock relativistic hadronic pressure then gives a kick to the piled up
matter the result of which is the ejection of outflow from the shock surface.
For this type of inflow, accretion is known to proceed smoothly
after a shock transition, since successful subsonic solutions
have been constructed (Flammang, 1982) for accretion onto black holes
embedded within normal stars with the boundary condition
${u = c}$; where $u$ is the infall velocity of matter and $c$ is the
velocity of light in vaccum.
The fraction of energy converted, the shock compression ratio
${R_{comp}}$ (in the notation of DC98),
along with the ratio of post shock relativistic
hadronic pressure to infalling ram pressure at a given shock location
are obtained from the steady state shock solution of 
Ellision and Eichler (Ellison \& Eichler, 1984;1985).
The shock location as a function of the specific energy ${\cal E}$ of the
infalling matter and accretion rate is then self consistently obtained
using the above mentioned quantities.
We then calculate the amount of mass outflow rate $R_{\dot m}$ from the 
shock surface using combination of {\it exact} transonic inflow outflow
solutions and study the dependence of $R_{\dot m}$ on various physical
entities governing the inflow-outflow system.
We thus quantitatively compute the mass outflow rate {\it only } from the 
inflow parameters. In this way we analytically connect the accretion and
wind type topologies self-consistently. As there was no such attempt
available in the litereture which computes the mass loss rate from 
zero angular momentum quasi-spherical Bondi type accretion,
our work, for the {\it first time} we believe, could shed light on the 
nature of the outflow from the models of AGNs with no accretion disks. \\
We consider polytropic inflow. The outflow is also
assumed to be polytropic except the fact that ${{\gamma}_{outflow}}$
is assumed to be
less than the ${{\gamma}_{inflow}}$ reason of which 
will be discussed in the next section.
As a fraction of the energy of infalling material is converted into
radiation, energy flux of the wind is somewhat less than that of the accretion
but is kept constant througout the outflow.\\
The plan of the paper is as follows :\\
In the next Section, we describe
our model and present the governing equations for the inflow and the outflow
along with the solution procedure. In \S 3, we present the result of 
our computation. Finally, in \S 4, we draw our conclusions.
A preliminary report of this work has been presented elsewhere (Das, 1998b).\\
\noindent
\section{Model description, governing equations and the solution procedure}
\noindent
\subsection{Inflow model}
\noindent
We assume that a Schwarzschild type black hole of mass 10${M_{\odot}}$
quasi-spherically accretes low energy (${\cal E} \lsim 0.001$) fluid
obeying polytropic equation of state. The density of the fluid is $\rho(r)$, 
r being the radial distance measured in the 
unit of Schwarzschild radius. We also assume that the accretion rate (in the unit of
Eddington rate) with which the fluid is being accreted, is not a function
of $r$. For simplicity of calculation, we choose geometric unit 
(unit of length = $\frac{2GM}{c^2}$ $\&$ unit of velocity is $c$.
$M$ is the mass of the black hole, $c$ is the velocity of light and
$G$ is the universal gravitational constant. All other relevant 
physical quantities can be expressed like wise)
to measure all the relevant quantities. We ignore the 
self gravity of the flow and the calculation is being done using
Paczy$n^{\prime}$ski-Wiita (Paczy$n^{\prime}$ski \& Wiita, 1980)
potential which mimics surronding of the Schwarzschild 
black hole. The equations
(in dimentionless geometric unit) governing the inflow are :\\
a) Conservation of specific energy is given by:
$$
{\cal E}=\frac{{u(r)}^2}{2} + n{{a(r)}^2} -\frac{1}{2(r-1)}
\eqno{(1)}
$$
$u(r)$ and $a(r)$ are the radial and polytropic sound velocities respectively.
$a(r)=\left[\frac{{\gamma}p(r)}{{\rho}(r)}\right]^{\frac{1}{2}}$;  $p(r)$ being the polytropic
pressure. For a polytropic inflow, 
$p(r)=K{{\rho(r)}^{\gamma}}$ where K is a measure of the entropy which remains
constant throught the flow except at the shock location. $n$ is the 
polytropic constant of the inflow $n={(\gamma - 1)^{-1}}$, $\gamma$ 
being the polytropic index.\\
b) Mass conservation equation is given by,
$$
{{\dot M}_{in}}={{\Theta}_{in}}{\rho(r)}u(r){r^2}
\eqno{(2)}
$$
As already mentioned, we assume that a steady, collisionless shock forms
at a distance $r_{sh}$ (measured in the unit of Schwarzschild radius) due
to the instabilities in the plasma flow. We also assume that for our model,
the effective thickness of the shock $\Delta_{sh}$ is small enough compared
to the shock standoff distance, i, e,\\
$$
\Delta_{sh} << r_{sh}
$$
Accreting particles with infall velocity $u(r)$ are then assumed to be
shock accelerated via first order Fermi acceleration. Due to
this process, relativistic protons will be produced. These relativistic protons,
suffer essentially no Compton loss and hence are not readily 
swallowed by the black hole. Rather they usually scatter several times 
before being captured by the black hole thus provide sufficient 
outward pressure necessary to support the shock. These protons, in turn,
produce pions through post-shock inelastic nuclear collisions
$$
p + p \longrightarrow  p + p + \pi^{\pm} +  \pi^0
$$
Pions generated by this process, decay into relativistic electrons,
neutrinos $\&$ antinutrinos and
produces high energy $\gamma$ rays.\\
$$
\pi^0 \longrightarrow 2\gamma
$$
$$
\pi^{-} \longrightarrow e^{-} + \overline {\nu_e}
$$
$$
\pi^{+} \longrightarrow e^{+} + \nu_e
$$
These electrons produce the observed non-thermal radiation by Synchrotron
and inverse Compton scattering. The 
overall efficiency of this mechanism depends largly on the shock location.
It has been shown (Eichler, 1979) that almost half of the energy flux 
that goes into relativistic particles is lost due to neutrinos.\\
From eqs. (2), the density of infalling matter $\rho(r)$ comes out to
be
$$
\rho(r) = \frac{{\dot M}_{in}}{\Theta_{in}u(r)r^2}
\eqno{(3)}
$$
At the shock, density of matter will shoot up and inflow velocity fall
down abruptly. If $({\rho}_{-}, u_{-})$ and $({\rho}_{+}, u_{+})$ are the pre
and post-shock densities and velocities respectively, then
$$
\frac{{\rho}_{+}}{{\rho}_{-}} = R_{comp} = \frac{u_{-}}{u_{+}}
\eqno {(4)}
$$
Where $R_{comp}$ is the shock compression ratio (in the notation of DC98).
For high shock much number solution (which is compatible with our 
low energy accretion model), the expression for $R_{comp}$ 
can be well approximated as 
$$
R_{comp} = 1.44{M_{sh}}^{\frac{3}{4}}\\
\eqno{(5)}
$$
Where $M_{sh}$ is the shock Mach number and eqs. (5) holds good for
$M_{sh} \gsim 4.0$ (Ellison \& Eichler, 1985).
However, the scattering mean free path of the relativistic hardrons
produced by this process are assumed to be small enough so that 
they could encounter the full shock compression ratio while crossing the
shock.\\
The hadronic interaction charecteristic time scale $\tau_{pp}$ may be
expressed as
\footnote {As only a fraction of the accreting matter is 
shock enerzyzed, the value of $\rho_{+}$ used to calculate
$\tau_{pp}$ is, in reality, less than that of the actual
$\rho_{+}$, giving a higher value of $\tau_{pp}$. The accurate
value of $\tau_{pp}$ can be calculated using the cosmic ray energy spectrum
and coupling the relativistic
and non-relativistic part of the accreting plasma which will 
be presented elsewhere. Nevertheless, our rough estimation
assures that even with the accurate value of $\tau_{pp}$ (which
is higher than that of used here), the conditions of shock formation
are highly satisfied.}
$$
\tau_{pp} = \frac{1}{{{\rho}_{+}}{{\sigma}_{pp}}c}
\eqno{(6)}
$$
Where ${\sigma}_{pp}$ is the collision cross section for relativistic
protons. 
Using eqs. (4),  $\tau_{pp}$ can also be expressed as
$$
\tau_{pp} = \frac{1}{R_{comp}{{\rho}_-}{\sigma_{pp}}c}
\eqno{(6a)}
$$
If we now assume that a fraction ${\cal E}_F$ of the infalling energy is
converted into radiation through the hadronic collision ($p$ - $p$)
and mesonic $(\pi^\pm,\pi^0)$ decay, this ${\cal E}_F$ will allow
convergent steady state solutions and in the case of quasi-spherical
infall, allows the development and maintenance of a standing,
collisionless shock at a fixed distance $r_{sh}$ measured in the
unit of Scwarzschild radius. \\
The luminosity obtained from this energy ${\cal E}_F$ at the shock location
$r_{sh}$ is assumed to be ${\cal L}_1$. For a spherical shock surface
and taking care of neutrino losses,
the expression for ${\cal L}_1$ can be written as (KE86)
$$
{\cal L}_1 = 2\pi{{r^2}_{sh}}{{\rho}_{sh}}m_p{{u^3}_{sh}}{{\cal E}_F}
\eqno{(7a)}
$$
where $m_p$ is the mass of the proton and subscript $sh$ indicates that
the respective quatities are measured at the shock location.\\
If we assume that the pressure of the relativistic particles
${\cal P}_{rel}$ (uniform inside the shock), the average energy
density is then $3{\cal P}_{rel}$, so alternatively the luminosity
can be expressed in terms of the volume integral of the
the emissivity $\epsilon$ due to the hadronic ($p - p$) collition
where 
$$
\epsilon = \frac{3{\cal P}_{rel}}{\tau_{pp}}
$$
Thus the alternative expresion for the luminosity obtained as function
of ${\cal P}_{rel}$ would be,\\
$$
{\cal L}_2 = \frac{4\pi{{r^3}_{sh}}{\cal P}_{rel}}{\tau_{pp}}
\eqno{(7b)}
$$
Defining $\delta$ as the ratio of downstream relativistic particle 
pressure to incoming ram pressure at the shock, we obtain\\
$$
{\cal P}_{rel} = \delta{{\rho}_{sh}}m_p{{u^2}_{sh}}
\eqno{(8)}
$$
Equating (7a) and (7b) and putting the values of $\rho, {\cal P}_{rel},$
and $\tau_{pp}$ from eqs. (3), (6a) and (8) respectively, we get
the expression for the shock location as a function of various
inflow parameters as 
$$
r_{sh} = \frac{3{\sigma_{pp}}c{\dot M}_{in}}{{u_{sh}}^2{\Theta_{in}}}\left(\frac{\delta}{{\cal E}_F}\right)
\eqno{(9)}
$$
The ratio $\left(\frac{\delta}{{\cal E}_F}\right)$ as a function 
of shock Mach number $M_{sh}$ for a high shock Mach number solution
(low energy inflow) is obtained from the empirical solution deduced
by Ellison and Eichler (Ellison \& Eichler, 1984), after suitable
modification required for our model.\\ 

\subsection{Outflow model}
\noindent
In ordinary stellar mass loss computations (Tarafder, 1988 and references therein)
the outflow is assumed 
to be isothermal till the sonic point. This assumption is probably justified, since 
copious photons from the stellar atmosphere deposit momenta on the slowly outgoing 
and expanding outflow and possibly make the flow close to the isothermal. This need
not be the case for outflow from black hole candidates. Our {\it effective boundary layer},
being pretty close to the black hole, are very hot 
and most of the photons emitted 
may be swallowed by the black hole itself instead of coming out of the region and
depositing momentum onto the outflow. Thus, the outflow could be cooler than 
the isothermal flow in our case. We choose polytropic outflow with a different 
polytropic index ${{\gamma}_o} < {\gamma}$ due to momentum deposition. As a fraction
of infalling energy density (${\cal E}_F$) is converted into radiation, specific 
energy of the outflow is somewhat less than that of the inflow. Nevertheless,
the outflow specific energy is also kept constant throughout the flow.\\
The following two conservation laws are valid for the outflow :
$$
{\cal E}^{\prime} = \frac{v(r)^2}{2} + {n^{\prime}}{a(r)^2} - \frac{1}{2(r - 1)}
\eqno {(10)}
$$
$$
{{\dot M}_{out}} = {{\Theta}_{out}}{\rho(r)}v(r){r^2}
\eqno{(11)}
$$
Where ${\cal E}^{\prime}$ is the specific energy of the outflow and ${\cal E}^{\prime} <
{\cal E}$  and $n^{\prime} =
{({\gamma}_o - 1)^{-1}}$ is the polytropic constant of the outflow. ${{\Theta}_{out}}$
is the solid angle subtended by the outflow and $v(r)$ is the velocity of the outflow.\\
For simplicity of calculation, we assume that the outflow is also
quasi-spherical and
${{\Theta}_{out}} {\approx} {{\Theta}_{in}}$. 
Defining $R_{\dot m}$ as the mass outflow rate, we obtain
$$
{R}_{\dot m} = \frac{{\dot M}_{out}}{{\dot M}_{in}}
\eqno{(12)}
$$
It is obvious from the above discussion 
that ${R}_{\dot m}$ should have some complicated functional dependences on 
the following parameters
$$
{R}_{\dot m} = {\Psi}({\cal E},{{\dot M}_{in}},{r_{sh}},{R_{comp}},{\gamma},{{\gamma}_o})
\eqno{(12a)}
$$
\subsection{Procedure to solve the inflow and outflow equations simultaniously}
\noindent
Before we proceed into further detail, a general understanding of the transonic
inflow outflow system in present case is essential to understand the basic scheme of the 
solution procedure. Let us consider the transonic accretion first. Infalling matter
becomes supersonic after crossing a saddle type sonic point, location of which is
determined by the inflow parameters such as the specific energy ${\cal E}$, ${\dot M}_{in}$
(in the unit of Eddington rate) and $\gamma$ of the {\it inflow}. This supersonic flow then
encounters a shock (if present) location of which ($r_{sh}$) is determined from eqs. 9.
At the shock, part of the incoming matter, having higher entropy density, is likely
to return back as wind through a sonic point other than the one 
through which it just entered. 
Thus a combination of transonic topologies, 
one for the inflow and other for the outflow
(passing through a different sonic point and following the topology {\it completely
different} than that of the ``self-wind" of the accretion), is required to obtain a full
solution. So it turns out that finding a complete set of self-consistent
inflow outflow solutions ultimately boils down to locating 
the sonic point of the {\it outflow} and the mass 
flux through it. Thus a supply of parameters ${\cal E}, {{\dot M}_{in}}, {\gamma}$ and 
${\gamma_o}$ make a self consistent computation of ${R_{\dot m}}$ possible.
Here $\gamma_o$ is supplied as free parameter because the self-consistent computation
of $\gamma_o$ directly using ${\cal E}, {{\dot M}_{in}}$ and $\gamma$ has not been attempted in this work,
instead we put a constrain that $\gamma_o < \gamma$ always and for any value of $\gamma$.
In reality, $\gamma_o$ is directly related to the heating and cooling procedure 
of the outflow.\\
The following procedure is adopted to obtain a complete solution: \\

From eqs. (1) $\&$ (2), we get the expression for the derivative
$$
\frac{du(r)}{dr} = \frac{\frac{2a(r)^2}{r} - \frac{1}{2(r - 1)^2}}{u(r) - \frac{a(r)^2}{u(r)}}
\eqno{(13)}
$$
At the sonic point, the numerator and denominator separately vanish and give rise to the
so called sonic point condition;
$$
u_c = a_c = \frac{\sqrt{r_c}}{2({r_c} - 1)}
\eqno{(14)}
$$
Where the subscripts $c$ represents the quantities at the sonic point. The derivative at the 
sonic point ${\left(\frac{du}{dr}\right)}_c$ is computed using the L' Hospital rule. 
The expression for ${\left(\frac{du}{dr}\right)}_c$ is obtained by solving the 
following polynomial,
$$
\left(\frac{2n+1}{n}\right){\left(\frac{du}{dr}\right)^2_c}
+{\frac{3u_c}{nr_c}}{\left(\frac{du}{dr}\right)}_c
-{\left\{{\frac{1}{(r_c-1)^3}}-\frac{2{a_c}^2}{r_c}\left(\frac{n+1}{n}\right)\right\}}
=0
\eqno{(15)}
$$

Using fourth order
Runge Kutta method, $u(r)$ and $a(r)$ are computed along the inflow from the {\it inflow} sonic point till
the position where the shock forms. The shock location is calculated 
by simultaniously solving the eqs.
(1), (2) and (9). With the known value of ${\cal E}^{\prime}$
and $\gamma_o$, it is easy to compute the location of the sonic point of the {\it outflow}
from eqs. (10) and (11). At the 
outflow sonic point, the outflow velocity $v_c$ and polytropic sound velocity $a_c$ is computed
in the same manner performed for the inflow. Using eqs. (10) and (11), $\left(\frac{dv}{dr}\right)$
and $\left(\frac{dv}{dr}\right)_c$ is computed as was done for the inflow. Runge-Kutta method is then
employed to integrate from the {\it outflow} sonic point towards the black hole to
find out the outflow velocity $v$ and density $\rho$ {\it at the shock location}.
The outflow rate is then computed using eqs.(12).\\
\section{Results}
Fig. 1 shows a typical solution which combines the accretion and the
outflow.  The input parameters are ${\cal E}$ = 0.001, ${\dot M}_{in}$
=1.0 Eddington rate (${\cal E}_d$ stands for the Eddington rate
in the figure)  and $\gamma = \frac{4}{3}$ 
corresponding to relativistic inflow.
The solid curve with an arrow represents the pre-shock region of the 
inflow and the solid vertical line with double arrow at $X_{pps}$ 
(the subscript $pps$ stands for $p$air $p$lasma mediated
$s$hock) represents
the shock transition. Location of shock is obtained using the eqs. $4$ 
for a particular set if inflow parameters mentioned above. Three
dotted curves show the three
different outflow branches corresponding to different polytropic 
index of the outflow as $\gamma_o$ = 1.3(left most curve),
1.275 (middle curve) and 1.25(rightmost curve).
It is evident from the figure that the outflow moves along the 
solution curves completely different from that of the 
``wind solution" (solid line marked with an outword directed arrow)
of the inflow which passes through the sonic point $P_s$. The mass loss
ratio $R_{\dot m}$ for these cases are 0.0023, 0.00065 and 0.00014 
respectively.
In Fig. 2., we have plotted the variation of $R_{\dot m}$ with 
incoming specific energy ${\cal E}$ for a set of values of
${\dot M}_{in}$ (measured in the unit of Eddington rate shown as ${\cal E}_d$ in the Fig.) shown in the 
Fig. It is observed that $R_{\dot m}$ monotonicaly increases with energy.
This is because as ${\cal E}$ increases keeping the Eddington rate of the inflow fixed,
the shock Mach number $M_{sh}$ decreases result of which is the 
decreament of shock location $r_{sh}$ and 
post shock density (via eqs. 5.) but the increament of the post 
shock fluid velocity ($v_{sh}$) with which the matter leaves 
the shock surface. The outfow rate $R_{\dot m}$, which is the product of
these three quantities, 
in general increases monotonically 
with ${\cal E}$ due to the combined tug of war of these three quantities.
Morever, closer the shock forms to the black hole, the greater will be 
the amount of gravitational potential available to be put onto
the relativistic hadrons
to provide more outward
pressure at the shock boundary which gives a stronger ``kick" to the 
accreting matter, the  result of which is the increament in $R_{\dot m}$.
All these points are manifested in Fig. 3 where we have 
shown the
variation of $R_{\dot m}$ as a function of 
compression ratio $R_{comp}$ (solid curve),
the shock location $r_{sh}$ (dotted curve) and the
injection velocity of the outflow $v_{sh}$ (dashed curve). The fig.
is drawn for a fixed $\gamma = \frac{4}{3}$ and ${{\gamma}_o}$ = 1.3
$R_{comp}$ and $v_{rsh}$ are scaled as ${R_{comp}}$ $\rightarrow$
($R_{comp} - 5.890)\times10^3$ and $v_{rsh}$ $\rightarrow 4\times10^{-6}v_{rsh}$.
The unequal gaps between the curves with different ${\cal E}_d$ 
implies that when inflow energy ${\cal E}$ is kept fixed,
$R_{\dot m}$ nonlinearly increases with Eddington rate.
This is because, as ${\cal E}$ is kept fixed while ${\dot M}_{in}$
is varied, the amount of infalling energy converted to produce
high energy protons is also fixed, so higher is the value of 
${\dot M}_{in}$ (in the unit of Eddington rate), the larger is the
distance of the shock surface (measured from the black hole) and the
outflowing matter feels low inward gravitational pull, the result of which
is the non-linear corelation of $R_{\dot m}$ with ${\dot M}_{in}$.\\
To have better insight of the behavior of the outflow, 
in Fig. 4a. and 4b, we plot $R_{\dot m}$ as a function of the 
polytropic index of the inflow $\gamma$ (Fig. 4a) and that 
of the outflow $\gamma_o$ (Fig. 4b) for fixed ${\cal E}$ = 0.001
and ${\dot M}_{in}$ = 1.0${\cal E}_d$. The range of $\gamma$ 
shown here are the range for which shock forms for the specified
${\cal E}$ and ${\dot M}_{in}$. In Fig. 4b, it is observed that shock
forms for the values of $\gamma_o$ lower than that shown in the
figure, but mass loss rate is then so small that we did not plot for
$\gamma_o < 1.23$. The general conclusion is that
$R_{\dot m}$ correlates with $\gamma_o$. This is because as $\gamma_o$
increases, shock location and post shock density of matter does 
not change (as $\gamma_o$ does not have any
role in shock formation or in determining the $R_{comp}$) but the sonic point of the {\it outflow}
is pushed inward, hence  the velocity with which outflow leaves the shock surface
goes up resulting the increament in $R_{\dot m}$. However, $R_{\dot m}$ anticorelates with $\gamma$
which is observed from Fig. 4a.\\
\section{Concluding Remarks}
In the present paper, we have computed the mass outflow rate from the
relativistis low energy matter quasi-spherically accreting onto
Schwarzschild type black holes. The free parameters chosen for the inflow
are the specific energy ${\cal E}$, mass inflow rate ${\dot M}_{in}$ (In
the unit of Eddington rate) and the polytropic index $\gamma$ of the inflow.
Only one extra parameter which was supplied for the outflow is its polytropic
index $\gamma_o$. We have computed the mass outflow rate $R_{\dot m}$ taking the
help of inflow parameters only (except $\gamma_o$) thus could analytically connect the 
accretion and wind type topologies self-consistently. In our computation,
we could investigate the dependence of $R_{\dot m}$ as a function of different
physical quantities governing the inflow.\\
The basic conclusions of this paper are the followings,
\begin{enumerate}
\item It is possible that outflows for quasi-spherical
Bondi type accretion onto a Schwarzschild black hole are coming from the
pair plasma pressure mediated shock surface.
\item The outflow rate monotonically increases with the specific energy of the 
inflow and nonlinearly increases with the Eddington rate of the infalling matter.
\item $R_{\dot m}$, in general, corelates with $\gamma_o$ but anticorelates
with $\gamma$.
\item Generaly speaking, as our model deals with high shock Mach number (low
energy accretion) solutions, outflows in our work always generate from the 
supersonic branch of the inflow, i,e, shock is always located {\it inside}
the sonic point.
\item Unlike the mass outflow from the {\it accretion disks} 
around black holes (Das, 1998; 1998a; DC98)
here we found that the value of $R_{\dot m}$ is distinguishably small. This is
because matter is ejected out due to the pressure of the relativistic 
plasma pairs which is {\it less enough} in comparison to the pressure
generated due to the presence of significant angular momentum.
However, in the present work we have dealt only high Mach number
solution which means matter is accreting with very low energy (cold inflow,
as it is described in litereture). This is another possible reason to 
obtain a low mass loss rate. If, instead of high Mach number solution,
we would use low Mach number solution, e,g, high energy accretion, the 
mass outflow would be considerably higher (this is obvious because
it has already been established in present work that $R_{\dot m}$
increases with ${\cal E}$. In our next work, we will present this 
type of model by calculating $R_{comp}$ and $\left(\frac{\delta}{{\cal E}_F}\right)$
for low Mach number solution.
\end{enumerate}
In the literature, we did not find any numerical simulation work which deal with
the type of outflow we discuss in this paper (outflow from zero angular
momentum inflow). On the other hand, observationally it is very difficult to
calculate the outflow rate exactly from a real system, as it depends on too
many uncertainities, such as filling factors and projection effects. So, at
present, we don't have any offhand results in this field with which our 
result could be compared.\\

There are a number of possible improvements which could be made on this 
preliminary work. For instance, the effect of radiation pressure on both
the inflow and the outflow could be taken into account. A preliminary
investigation shows that the effect of radiation force, when included
into the basic conservation equations (eqs. (1), (2) and (10), (11)) decreases
the value of $R_{\dot m}$. This is probably because introduction of any radiation
term which is proportional to $\frac{1}{r^2}$, wekens gravity and 
pushes the shock location outwards, combined effect of which is the decreament
of $R_{\dot m}$. Another possible improvement is to include the magnetic field
to give the outflow an appropiate geometry. In our model, we assumed the outflow
to be quasi-spherical as like the inflow. The introduction of magnetic field
probably could collimate the outflow thus would help to investigate 
the structure of jet related to the model of AGN 
without accretion disk in greater detail. Lastly, we did not self consistently compute $\gamma_o$
as a function of inflow parameters and $\gamma_o$ was
supplied as a free parameter. In our future work, we will be presenting the self
consistent calculation of $\gamma_o$ to reduce the number of free parameters
of the problem.\\

So far, we made the computation around Schwarzschild black hole. Our work
could be extended to study mass outflow in kerr space time using pseudo-Kerr
potential (Chakrabarti \& Khanna, 1992) and incorporating the frame
dragging effect. This is under preparation and will be presented elsewhere. \\[1cm]
\noindent
{\large\bf Acknowledgement :}\\
The author expresses his gratitude to Prof. S. K. Chakrabarti  and his colligue
Mr. A. Ray for constructive discussions and helpfull suggestions which 
improved the quality of this work.

\newpage
\begin{center}
\underline
{\large\bf Figure Captions}\\[1cm]
\end{center}
\noindent
Fig. 1: Solution topology for three different $\gamma_o$ (1.3, 1.275, 1.25)
for ${\cal E}$ = 0.001, ${\dot M}_{in}$ = 1.0, ${\cal E}_d$ = 1.0, $\gamma = \frac{4}{3}$.
$P_s$ indicates the sonic point of the {\it inflow} where $X_{pps}$ stands for the
shock location. See text for details.\\

\noindent
Fig. 2: Variation of $R_{\dot m}$ with {\it inflow} specific energy ${\cal E}$ for
different Eddington rate (marked as ${\cal E}_d$ in the figure). Other parameters
are $\gamma=\frac{4}{3}$ and $\gamma_o$=1.3. $R_{\dot m}$ monotonically increases
with ${\cal E}$, see text for detail.\\

\noindent
Fig. 3: Variation of $R_{\dot m}$ with the compresion ratio $R_{comp}$ (solid curve),
shock location $X_{ps}$ (dotted curve) and {\it outflow} velocity at shock
$v_{sh}$ (dashed curve). Other parameters are ${\cal E}$=0.001, ${\dot M}_{in}$= 1.0(In
the unit of Eddington rate), $\gamma=\frac{4}{3}$ and $\gamma_o$=1.3.\\

\noindent
Fig. 4a: Variation of $R_{\dot m}$ with polytropic index of {\it inflow} $\gamma$. Other
parameters are ${\cal E}$=0.001, ${\dot M}_{in}$=1.0 (in the
unit of Eddington rate) and $\gamma_o$=1.3.\\

\noindent
Fig. 4b: Variation of $R_{\dot m}$ with polytropic index of {\it outflow} $\gamma_o$. Other
parameters are ${\cal E}$=0.001, ${\dot M}_{in}$=1.0 (in the
unit of Eddington rate) and $\gamma=\frac{4}{3}$.
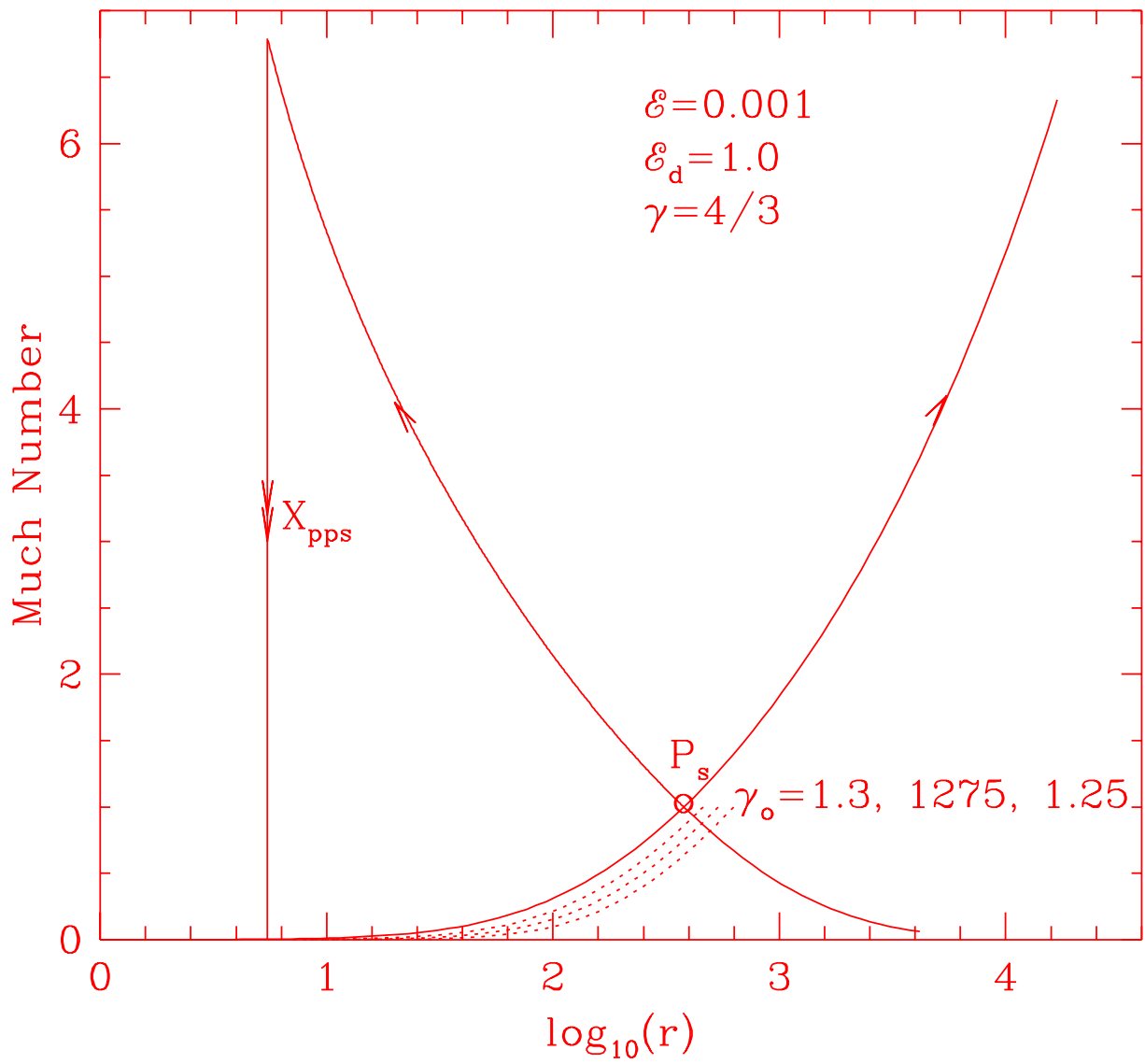
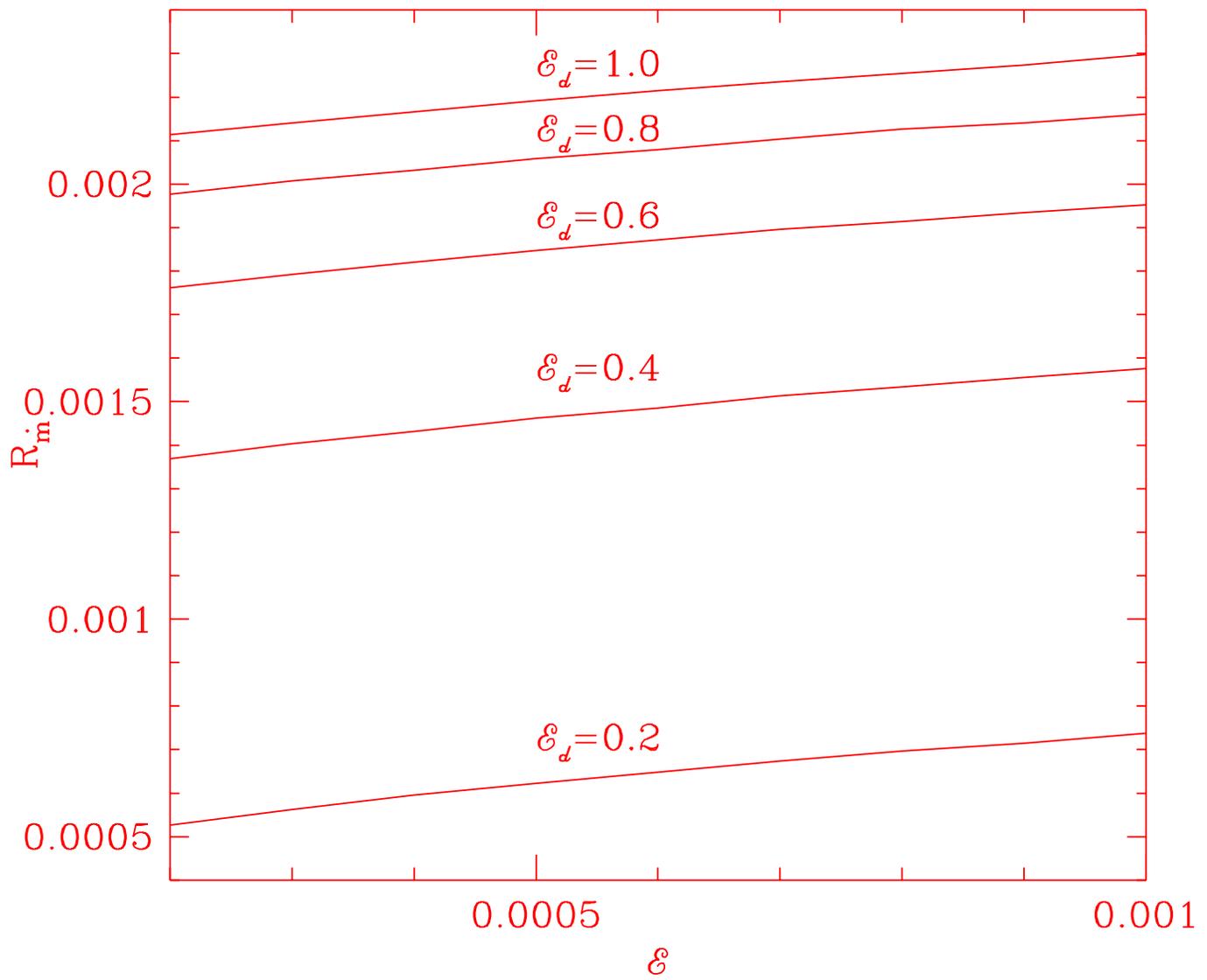
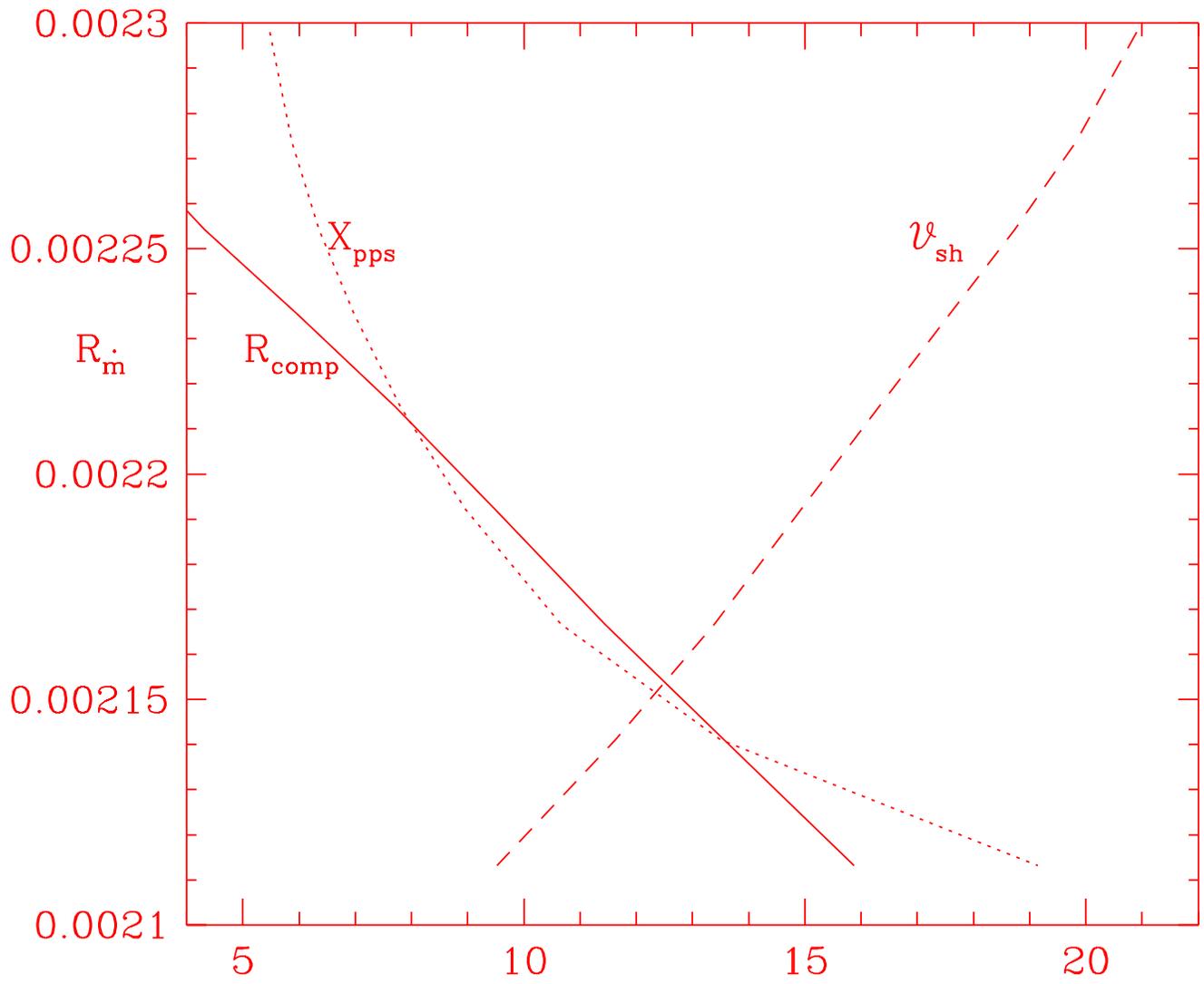
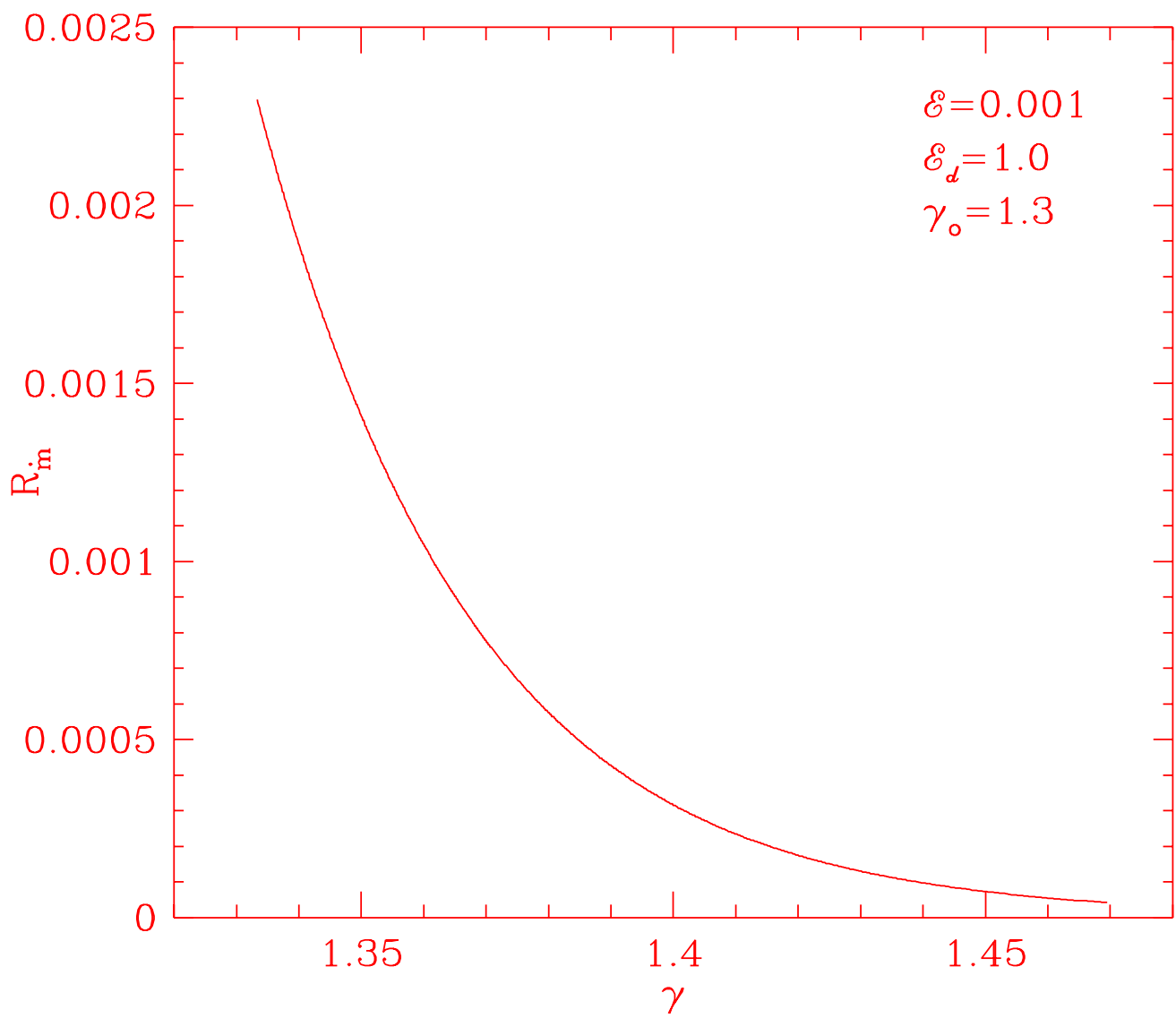
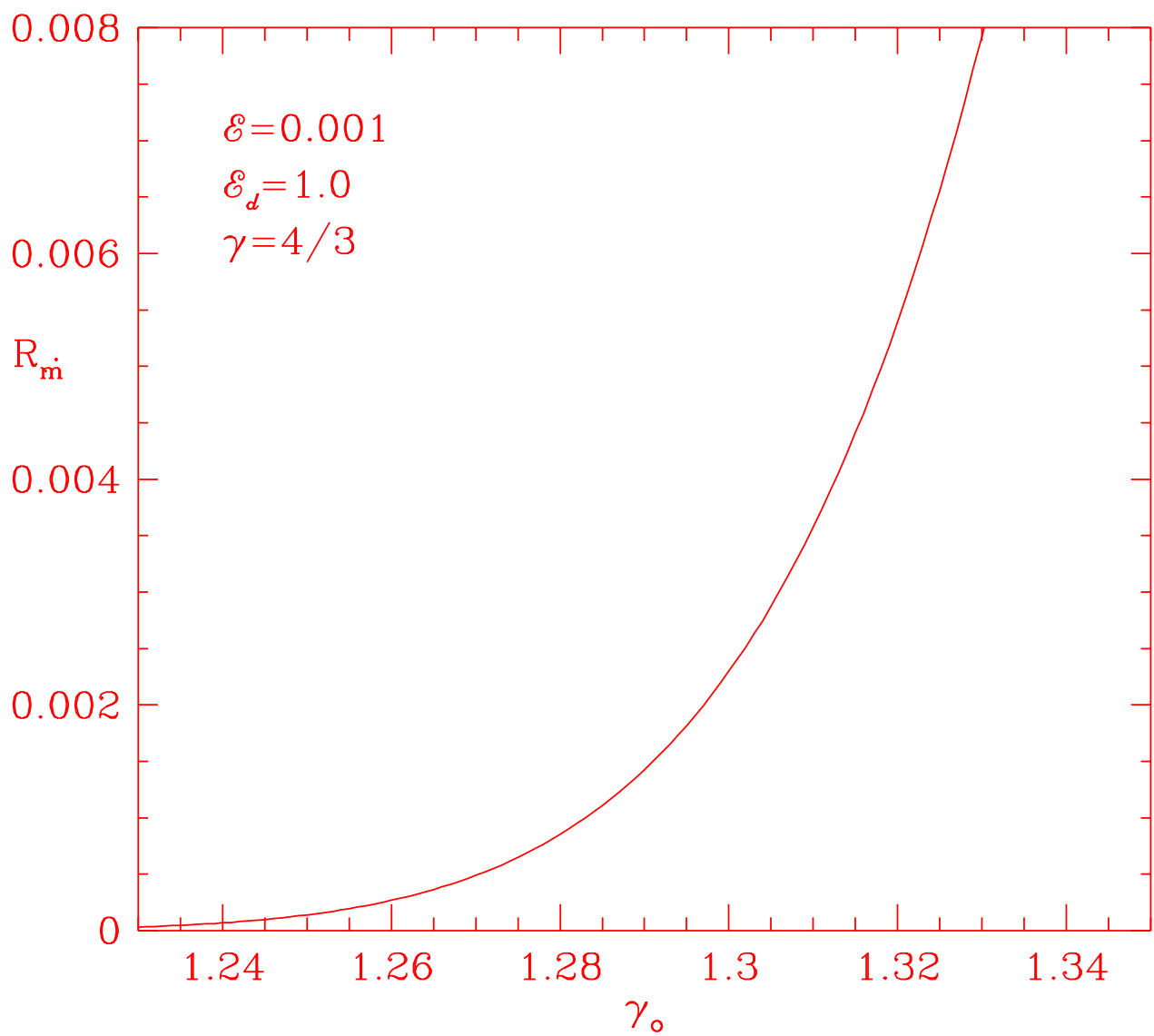
\end{document}